\documentclass[] {aipproc}

\layoutstyle{8x11single}

\SetInternalRegister\hbadness{8000} % pseudo latin isn't breaking very well :-)

\newcommand\doingARLO[2][]{%
  \ifx\mmref\undefined #1\else #2\fi
}

\begin{document}

\title{First results from RHIC: What are they telling us?}


\author{J.L. Nagle}{
  address={922 Pupin Hall, Columbia University, New York, NY 10027},
  email={nagle@nevis.columbia.edu},
  thanks={}
}

\copyrightyear  {2001}

\begin{abstract}
The Relativistic Heavy Ion Collider (RHIC) facility at Brookhaven National Laboratory
is the first accelerator specifically constructed for the study of very hot and dense
nuclear matter.  At sufficiently high temperature, nuclear matter is expected to undergo
a phase transition to a quark-gluon plasma.  It is the specific goal of the field to
study the nature of this plasma and understand the phase transitions between
different states.  The RHIC accelerator along with four experiments BRAHMS, PHENIX, PHOBOS, and
STAR were commissioned last year with first collisions occurring in June 2000.  Presented
here are the first results from low luminosity beam in Run I.  They are a glimpse of
the wealth of physics to be extracted from the RHIC program over the next several years.
\end{abstract}

\date{\today}

\maketitle

\section{Introduction}
The long awaited first results from the RHIC facility are now
available from all four experiments in the program.  Here we
review the fundamental physics goals of the field of Relativistic
Heavy Ions. We begin with a brief overview of the four experiments
and the conditions for Run I at RHIC.  Subsequently the early
physics results are presented in three sections including
understanding the initial conditions, the thermalization of the
system, and specific probes of the possible plasma phase. Finally
we present an outlook for Run II at RHIC, which is currently
underway, and the long term future of the program.

\section{Physics Goals}
The field of Relativistic Heavy Ions has one primary goal.  That
goal is to understand the behavior of QCD under extreme conditions
of high temperature and density.  In this limit, non-perturbative
QCD may simplify to the extent that it is possible to employ a
thermodynamic description of a quark-gluon plasma phase
characterized by a screening of the confining potential for
partons and the restoration of approximate chiral symmetry. The
nature of the QCD vacuum under these conditions is of considerable
interest. Calculations in the framework of Lattice QCD indicate
that in the low net baryon density limit there may be a phase
transition at a temperature of order 150-200 MeV. Currently
lattice calculations cannot determine the order of the transition
because of theoretical uncertainties associated with the number of
light quark flavors included (either two or three if including the
strange quark). Thus the order of the transition is currently an
experimental question.

Of secondary interest in the field is the implications of the QCD
phase transitions on the fields of cosmology and astrophysics.
Shortly after the Big Bang the universe is believed to have
existed in the quark-gluon plasma phase.  As the universe expanded
and cooled, the universe went through a hadronization phase
transition at a temperature of 150-200 MeV. Witten \cite{Witten}
and others \cite{Applegate} have speculated that a strong first
order phase transition in the early universe with associated
bubble formation could have led to large inhomogeneities in the
baryon density.  If the inhomogeneities survived the diffusion
process until a temperature of approximately 1 MeV, then they
could have observable consequences for Big Bang Nucleosynthesis.
There has been continuing work on the modeling of this bubble
nucleation and diffusion rates to address this question.  It would
be a great scientific achievement to determine the order of this
transition experimentally from Relativistic Heavy Ion Collisions.
In addition, quite recently many theorists have speculated on the
nature of dense quark matter, possibly a color superconductor,
at the core of neutron stars and
other compact objects \cite{Krishna}.

It is important to note that particularly at RHIC energies, there
are many cross-overs between the physics issues addressed in this
field and those in other fields such as: deep inelastic
scattering, hadron colliders, TJNAF structure function physics,
and condensed matter physics.  Hopefully this will broaden the
scope of the field and also lead to greater inter-field
collaboration.

\section{RHIC Accelerator and Experiments}
There are four experiments currently implemented for the RHIC heavy ion program, with
two additional intersection regions available for future experiments.
The two smaller experiments are BRAHMS and PHOBOS, and the two larger experiments
are PHENIX and STAR.

The BRAHMS experiment consists of two spectrometer arms that are
capable of rotating around a pivot in order to cover a large range
of rapidity and transverse momentum. The detectors used include
Time-Projection-Chambers (TPC), wire chambers, Cerenkov counters,
and a time-of-flight system.  Their experimental emphasis is
particle identification over a broad rapidity range.  They cover
nearly the full 12 units of rapidity, well beyond the other
experiments' capabilities.  This may allow them to explore the more
baryon dense matter at forward and backward rapidities.

The PHOBOS experiment has two main features.  The first is nearly
$4\pi$ coverage with silicon detectors to measure the total
charged particle multiplicity, enable phase space fluctuations
analysis, and measure global properties such as elliptic flow. The
second part consists of two spectrometer arms, also silicon based,
that give good particle identification over a smaller acceptance
for measuring two particle correlations and reconstructing the
$\phi$ in the $K^{+}K^{-}$ decay channel.  They also have partial
coverage with a time-of-flight wall to enhance particle
identification.  PHOBOS emphasizes high data rate and fluctuation
analysis capabilities.

The PHENIX experiment is specifically designed to measure
electrons, muons, hadrons and photons. The experiment is capable
of handling high event rates, up to ten times RHIC design
luminosity, in order to sample rare signals such as the $J/\psi$
decaying into muons and electrons, high transverse momentum
$\pi^{0}$'s, direct photons, and others.  The detector consists of
four spectrometer arms.  Two central arms have a small angular
coverage around central rapidity and consist of a silicon vertex
detector, drift chamber, pixel pad chamber, ring imaging Cerenkov
counter, a time-expansion chamber, time-of-flight and an
electromagnetic calorimeter.  These detectors allow for electron
identification over a broad range of momenta in order to measure
both low mass and high mass vector mesons. Two forward
spectrometers are used for the detection of muons. They employ
cathode strip chambers in a magnetic field and interleaved layers
of Iarocci tubes and steel for muon identification and triggering.

The principal detector in the STAR experiment is a large volume
Time-Projection-Chamber (TPC). The TPC is augmented with a silicon
vertex detector, a small coverage ring imaging Cherenkov (RICH)
counter, and an electromagnetic calorimeter. The experiment
emphasizes large phase space coverage and particle identification
via $\it{dE/dx}$, augmented by the limited coverage RICH and a
future additional Time-of-Flight wall.  Measurements of strange
particles including strange baryons $\Lambda, \Xi, \Omega$ and
antibaryons $\overline{\Lambda}, \overline{\Xi}, \overline{\Omega}$, two
particle correlations and event-by-event fluctuations are among
the many observables accessible to STAR. The large phase space
coverage of the experiment allows many critical measurements to be
done event-by-event, for example hadron spectra, two particle
correlations, and particle ratios.

The design of both the STAR and PHENIX experiments includes a
polarized proton program to conduct studies of the spin structure
of the proton.  Critical to this measurement is the identification
of high transverse momentum photons and leptons. The STAR
experiment is phasing in an electromagnetic calorimeter that will
be crucial for such observations.  In particular the RHIC
experiments are ideally suited for measuring the gluon
contribution of the proton's spin.

\section{Run I}
The RHIC facility is designed to accelerate fully stripped Au ions
to a collision center-of-mass energy of 200 GeV per nucleon pair.
The design luminosity is $2 \times 10^{26}$ ions/s/cm$^{2}$, which
corresponds to approximately 1400 Au + Au minimum bias collisions
per second.  During Run I at RHIC the maximum energy was 130 GeV
per nucleon pair and they achieved 10\% of design luminosity by
the end of the run. First collisions were achieved in June 2000
and the run continued until the end of August.  All four
experiments integrated sufficient luminosities to commission the
detectors and produce a variety of initial physics results. It
should be noted that the total statistics recorded, for example by
the PHENIX experiment, is equivalent to less than one day of
sampled collisions at full design luminosity in  Run II. Thus the 
results presented below are
just a preliminary look at the physics potential of the RHIC
experiments.

\section{Time Evolution of the Collision}
The time evolution of RHIC collisions can be broken down into
various stages.  The initial stage is that of two Lorentz
contracted Au nuclei incident upon each other at
ultra-relativistic velocities. This may seem like an uninteresting
stage, but there is a great deal of physics in the parton
distributions of the nuclei as described by nuclear structure
functions.  In deep inelastic scattering experiments the parton
density increases dramatically at low values of x (momentum
fraction carried by the parton relative to the nucleon's total
momentum).  Due to the non-Abelian nature of the strong
interaction, two gluons can fuse to form a single gluon.  Thus, at
small enough x, the gluon density may in fact saturate.  In
electron-proton experiments at HERA saturation effects are being
investigated at very low $x < 10^{-3}$.  However in a nucleus the
gluon wavefunction from many different nucleons can overlap and
thus saturation may occur at much higher values of $x \approx
10^{-2}$, which happens to be the $x$ scale relevant to collisions
at RHIC energies. The RHIC experiments have already begun to
address this question. This represents an interesting
non-traditional heavy ion physics topic and has large overlap with
structure function physics.  This has motivated some to propose an
electron beam to study eA collisions at RHIC (eRHIC), thereby
extending the range of the machine to very low $x$.

The next stage is that of pre-equilibrium.  In a time of less than
0.1 fm/c the two nuclei pass through each other and of order
10,000 quarks, antiquarks and gluons collide.  Some of these
collisions are at large momentum transfer and thus have
calculable rates using perturbative QCD.  For example, a high
transverse momentum quark represents a calculable probe that is
generated before any equilibrium is established, but must then
traverse the remaining system before fragmenting into a jet cone.

After a period of time of 0.1 to 1.0 fm/c the bath of semi-hard
quarks and gluons may equilibrate.  It is interested to note that
if parton saturation is achieved by the copious production of
gluons, the system essentially thermalizes immediately at 0.1
fm/c. What is the nature of the bath of quarks and gluons?  Since
they are at low relative momentum this can only be reliably
calculated in the framework of lattice QCD.  Lattice results
indicate that a phase transition from confined hadronic matter to
a quark-gluon plasma takes place for temperatures in excess of 150
- 200 MeV. In calculations including only two light quark flavors
the phase transition appears to be first order, while including
the strange quark as a ``light quark'' changes the transition to
second order. A key goal of the field is the determination of the
phase transition temperature and the order of the transition.

Since the system has no external constraints it expands and cools
off.  The analogy of the early universe expanding and cooling
after the Big Bang is quite appropriate.  If the system has
equilibrated then the expansion may be describable via
hydrodynamics and an appropriate equation of state. Once the
system cools below the critical temperature, the quark and gluons
hadronize.  The hadrons continue to scatter and interact until
eventually all interactions cease (freeze-out), after which
particles freely stream away from the collision point.  The point at which
inelastic collisions cease is referred to as chemical freeze-out, when 
all particle ratios are frozen.  The stage when all elastic
collisions cease is referred to as thermal freeze-out, when the
momentum distributions of the particles is frozen.
Measurements of hadron spectra and correlations give a photograph
of the system at the point of thermal freeze-out.

\section{Results}

Here we divide the early results into three sections detailing:  (1) the initial conditions,
(2) the thermodynamic properties, and (3) the probes of the plasma state.  It should be 
noted that this is not meant to be a comprehensive review of all results to date, but
just a select sample.  Many important and interesting results not included below can be
found in the recent Quark Matter 2001 conference proceedings~\cite{qm2001}.

\subsection{(1) Initial Conditions}

\begin{figure}
  \includegraphics[height=.3\textheight]{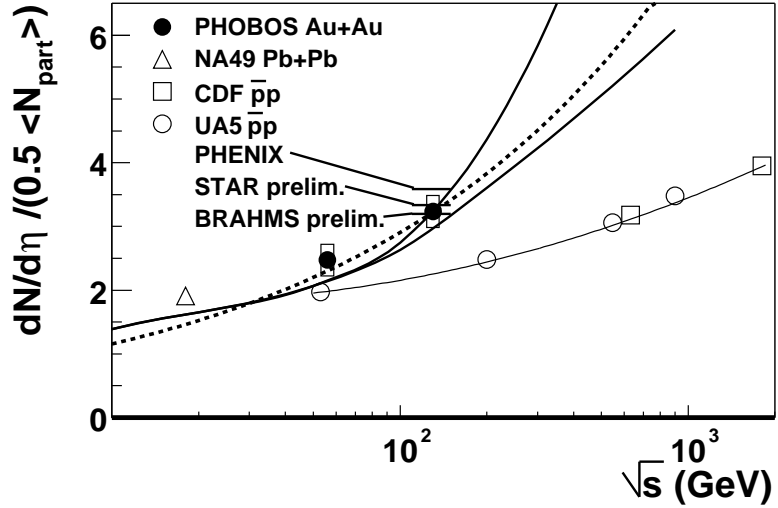}
  \caption{Charged particle multiplicity measurement from all four RHIC experiments is shown
for Au+Au collisions at $\sqrt{s_{NN}}=130~GeV$.  Also shown
are data for $p+p$ and $p+\overline{p}$ collisions.  A model of heavy ion collisions HIJING is
shown for comparison.}
\label{fig:charge}
\end{figure}

The first result with a measurement from all four RHIC experiments
is the charged particle multiplicity
\cite{phobos-prl1,phenix-nch,brahms-qm,star-qm}. The four experiments' results for
central (small impact parameter) collisions are in excellent
agreement and are shown in Fig.~\ref{fig:charge}.  The multiplicity 
rises more sharply as a function of center-of-mass energy in heavy ion
collisions than in $p+p$ and $p+\overline{p}$ collisions, which is
attributed to the increased probability for hard parton scattering in 
the thick nuclear target seen by each parton.

We expect the charge particle yield
to increase for collisions of larger nuclei.  However, at low x values, the 
high density of gluons may in fact saturate due to gluon fusion processes.  
The contribution to
the yield from hard processes should exhibit point-like scaling
(scaling with the number of binary collisions) and would thus
scale as $A^{4/3}$.  However, parton saturation depends upon the
nuclear size and would limit the growth of the number of produced
partons as $A^{1/3}$.  If present, this initial parton saturation
would limit the hard process contribution to the total charged
particle multiplicity.

\begin{figure}
  \includegraphics[height=.3\textheight]{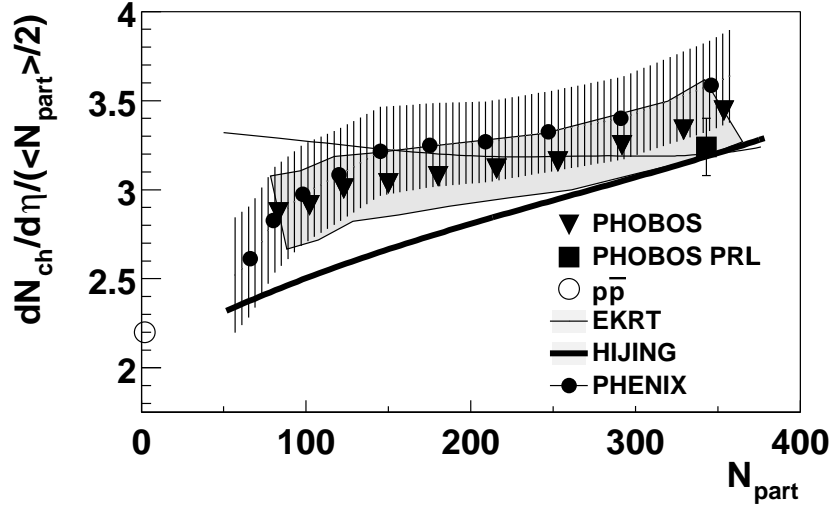}
  \caption{PHENIX and PHOBOS results for
$dN_{ch}/d\eta|_{\eta=0} / {1 \over 2}N_{part}$ as a function of $N_{part}$.
The hashed and solid bands indicate the systematic errors for the two
experimental results.  
The data point for $p\overline{p}$ with two participants is shown for
comparison.  Also theoretical predictions from the HIJING and EKRT models
are shown.}
\label{fig:npart}
\end{figure}

Only one nuclear species was accelerated in Run I at RHIC.  Thus,
rather that changing the  mass number $A$ directly we control the
collision volume by varying the centrality or the number of
participating nucleons for Au+Au collisions. Shown in
Fig.~\ref{fig:npart} are the published results from the PHENIX
\cite{phenix-nch} and PHOBOS \cite{phobos-nch2} experiments for
the number of charged particles per participant nucleon pair as a
function of the number of participating nucleons.   The number of 
participating nucleons is determined in a slightly different manner
by the different experiments.  However, the general method is to calibrate the
number of spectator nucleons (= $2 \times A$ - participant nucleons)
using a measurement of spectator neutrons in a set of zero degree
calorimeters that are common to all experiments.  By correlating the
number of forward neutrons to the number of charged particle produced in
the large pseudorapidity region, the event geometry can be understood.

In Fig.~\ref{fig:npart} one can also see
theory comparisons that indicate that a model including parton
saturation (EKRT~\cite{ekrt}) fails to agree with the more peripheral data.  Results
from the HIJING model~\cite{hijing} are also shown which does not include parton
saturation and thus has a more continuous rise in the particle
multiplicity.  Since
saturation phenomena are only likely to have observable
consequences for large collision volumes, it is not possible with
present systematics to rule out the saturation picture for the
most central collisions. 

In order to better test the saturation picture lighter ion, smaller $A$, 
collisions will be studied in Run II.  In addition, heavy flavor (charm and
bottom) and Drell-Yan production should be a sensitive probe to the
initial parton density.  Another proposal is that by varying the collision 
energy and keeping the nuclear geometry the same you can get a better
handle on systematics and test scenarios dependent on the coupling constant
and the saturation scale.

%Note charm and bottom might help solve this question.  Initial
%state with pA using Drell-Yan also might work.  Then go to eRHIC.
%Produced gluon saturation only in AA.  Experiment of varying the A
%for  the colliding system.  Maybe show Mueller's formula.

\begin{figure}
  \includegraphics[height=.3\textheight]{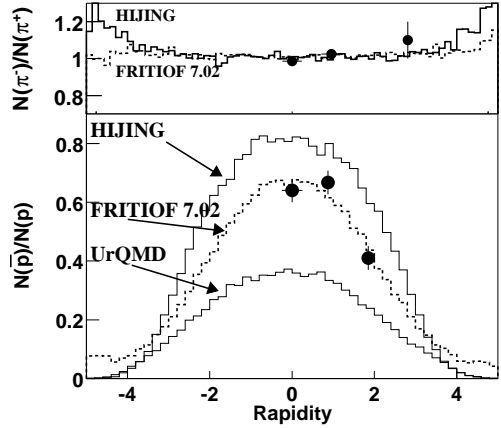}
  \caption{Plotted is the $\pi^{-}/\pi^{+}
$ and $\overline{p}/p$ ratio as a function of rapidity from the BRAHMS experiment 
for central collisions.}
\label{fig:brahms_pbarp}
\end{figure}

In addition to wanting to know the initial parton density, the energy density is of
great interest.  There are published results estimating the initial
thermalized energy density achieved in these collisions.
Bjorken originally derived a formula, shown in Eqn.~\ref{eqn:bj}, relating the measured
transverse energy per unit rapidity to the thermal energy density~\cite{bjorken}.
\begin{equation}
\epsilon_{B_{j}} = {{1} \over {\pi R^{2}}} {{1} \over {c\tau}} {{dE_{T}} \over {dy}}
\label{eqn:bj}
\end{equation}
It should be noted that there is a trivial
factor of two error in the original reference that is corrected here.
This formulation 
assumes a boost invariant expanding cylinder of dense nuclear
matter and a thermalization time $\tau$.  There are two important assumptions in this particular
formulation.  The first is the boost invariant nature of the collision.  
There are recent preliminary measurements from STAR and PHOBOS that
indicate the distribution of particles is relatively flat over $\pm 2$ units
of pseudorapidity.  However, shown in Fig.~\ref{fig:brahms_pbarp} is the
measured distribution of $\overline{p}/p$ from the BRAHMS experiment~\cite{brahms-ratio}.
This indicates the the system is already changing at rapidity $y\approx2$, though it
is not clear that this is enough to invalidate the energy density formulation.  
The second question is what is the relevant thermalization time $\tau$.

\begin{figure}
  \includegraphics[height=.3\textheight]{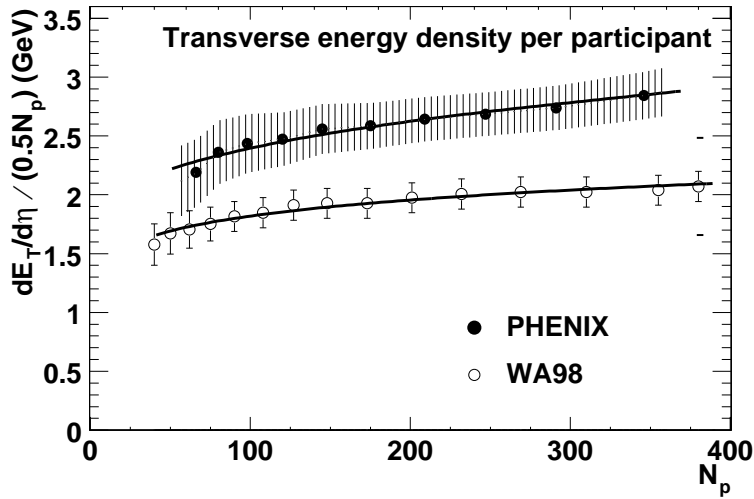}
  \caption{The PHENIX experiment result is shown for $dE_{T}/d\eta/(0.5N_{p})$ at $\eta=0$ as
a function of the number of participating nucleons.  Also shown in the 
result from experiment WA98 at the lower energy CERN-SPS.}
  \label{fig:phx-et}
\end{figure}

The PHENIX experiment has published \cite{phenix-et} the
transverse energy distribution for minimum bias Au+Au collisions.
For the 5\% most central events, the extracted transverse energy
$<dE_{T}/d\eta>|_{\eta=0} = 503 \pm 2$ GeV.  Shown in Fig.~\ref{fig:phx-et} 
is $dE_{T}/d\eta/(0.5N_{p})$ versus the number of participating
nucleons.  One sees a similar increase in transverse energy as was seen
in the charged particle multiplicity yield.  

The canonical
thermalization time used in most calculations is $\tau=1$ fm/c, that yields an energy density of
$4.6~GeV/fm^{3}$, which is is 60\% larger than measured at the
CERN-SPS.  In addition, it is believed that the density is
substantially higher due to the potentially much shorter
thermalization time in the higher parton density environment.  If
one achieves gluon saturation the formation time is of order 0.2
fm/c and gives an estimated energy density of $23.0~GeV/fm^{3}$.
There are even estimates of over $50~GeV/fm^{3}$, but they assume
a very large drop in the final measured transverse energy due to
work done in the longitudinal expansion of the system.
All of these estimates are above the energy density corresponding to the phase transition
temperature of 150-200 MeV is of order $0.6-1.8~GeV/fm^{3}$.

\subsection{(2) Thermodynamic Properties}

As the system approaches thermal and chemical equilibrium, we need
to have an experimental handle on the time scale and degree of
thermalization. The earliest and hottest stages of the collision that might
be characterized by deconfined matter will emit thermal radiation,
real photons $\gamma$ and virtual photons $\gamma^{*} \longrightarrow e^{+}e^{-}$ 
or $\mu^{+}\mu^{-}$, from quark-quark scattering.  
The resulting photons and leptons escape the collision
volume without re-interacting and thus act as a penetrating probe of the early
times.  This is analogous to measuring neutrinos from the center of the sun.
The data sample in Run I is not sufficient for these measurements, so higher
luminosity running in Run II is eagerly awaited.
However, the ratio of final hadron
yields has been measured and gives information on the equilibration condition at the later
stage of hadronic chemical freeze-out.  

\begin{figure}
  \includegraphics[height=.3\textheight]{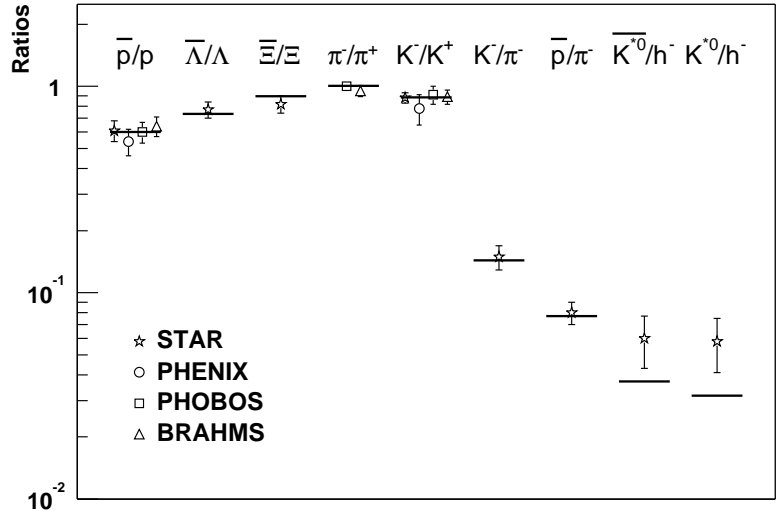}
  \caption{Particle ratios measured by all four experiments for Au+Au collisions.  
Note that many of these results are preliminary.}
\label{fig:ratios}
\end{figure}

All four experiments have shown particle ratios for
$\overline{p}/p \approx 0.6$ and are in excellent agreement with each other.
Both STAR and PHOBOS have measured other particle ratios including resonance
states such as the $K^{*}$ by STAR.  A collection of ratios (note that many are
preliminary results) are shown in Fig.~\ref{fig:ratios}.  These ratios can be
fit to a Grand Canonical Ensemble statistical model and show reasonable agreement
with a baryon chemical potential $\mu_{B} \approx 45-55$ MeV and a chemical
freeze-out temperature $T\approx 160-180$ MeV.  An example fit from ~\cite{pbm} is
shown as solid lines in Fig.\ref{fig:ratios}.  It should be noted that the preliminary
$K^{*}$ results are somewhat under predicted.  The measurement of other such excited
states should really test some of the model assumptions.
The freeze-out temperature is not
significantly different from that measured at the CERN-SPS, which is not
surprising since all systems must cool to approximately the same energy density
before particles stop having inelastic scatters.

Some have taken the ability to describe the system by this simple model as an
indication of thermalization.  However, a similar behaviour is seen in $e^{+}e^{-}$ and
$pp$, $p\overline{p}$ collisions with a limiting temperature $T\approx 170$ MeV, though
with an additional suppression of strangeness production.  There is most likely an
important connection between how a small region of vacuum with strong color fields 
fragments into hadronic states (for example in $p\overline{p}$ collisions) and how that 
occurs in a large volume as in heavy ion
collisions.  This connection needs to be further explored before any firm conclusion about
thermalization is made based on hadron ratios.

One of the most exciting early results from RHIC is the
measurement of elliptic flow.   For non-central collisions, the overlap
geometry between the two nuclei is almond shaped.  This spatial anisotropy
can translate into a momentum anisotropy in the presense of strong
re-scattering.  One can measure the resulting momentum anisotropy by 
measuring the second harmonic Fourier coefficient $v_2$ in the azimuthal distribution
of particles with respect to the reaction plane.  Since the flow is built
up from pressure gradients at the earliest stages of the time evolution, it may
be sensitive to initial large parton re-scattering in a deconfined phase and to a lesser
extent the Equation of State.

\begin{figure}[ht]
  \begin{minipage}[t]{0.49\textwidth}
    \includegraphics[width=1.1\textwidth]{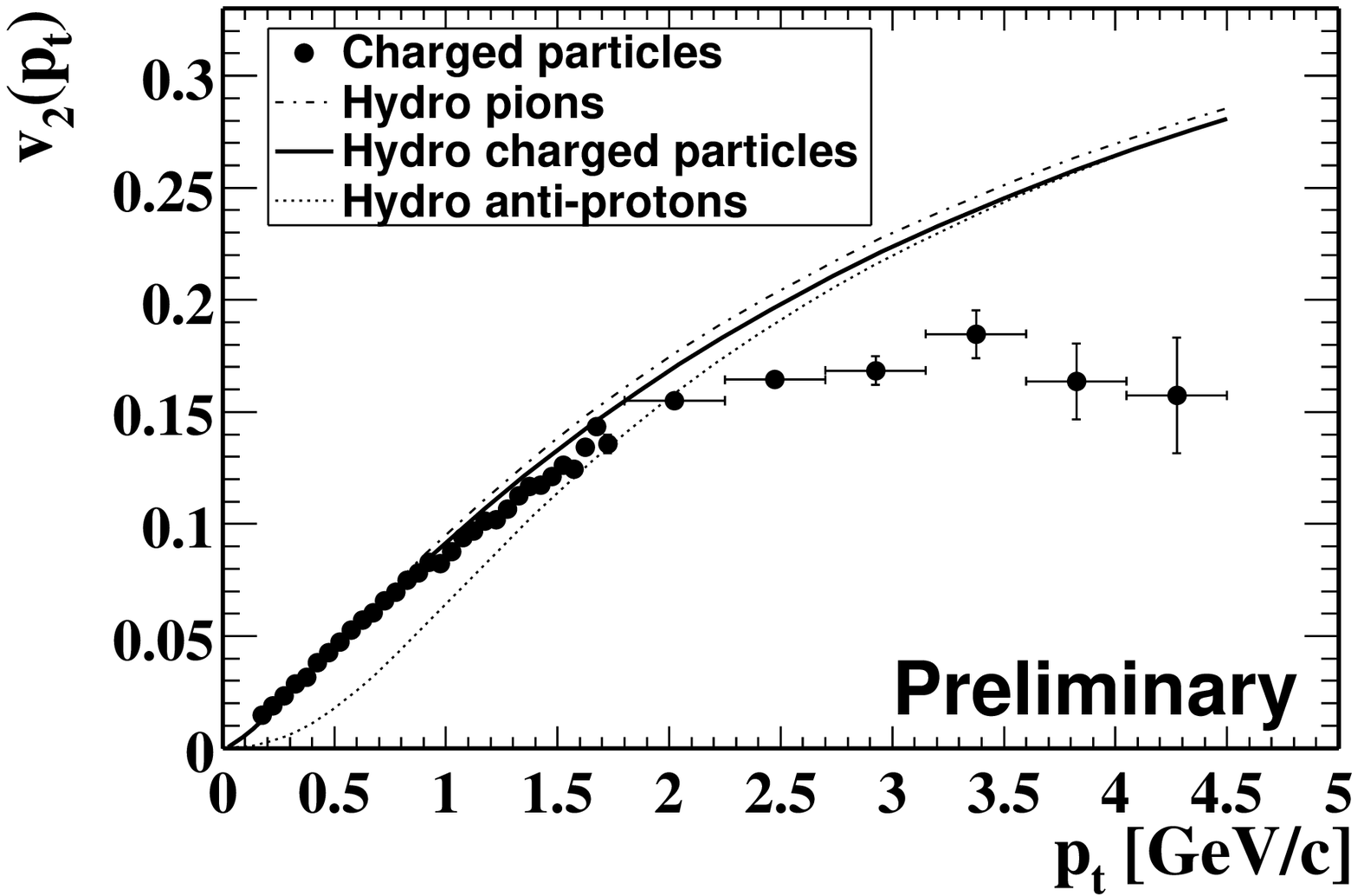}
    \label{highpt_hydro}
  \end{minipage}
  \hspace{\fill}
  \begin{minipage}[t]{0.49\textwidth}
    \includegraphics[width=1.1\textwidth]{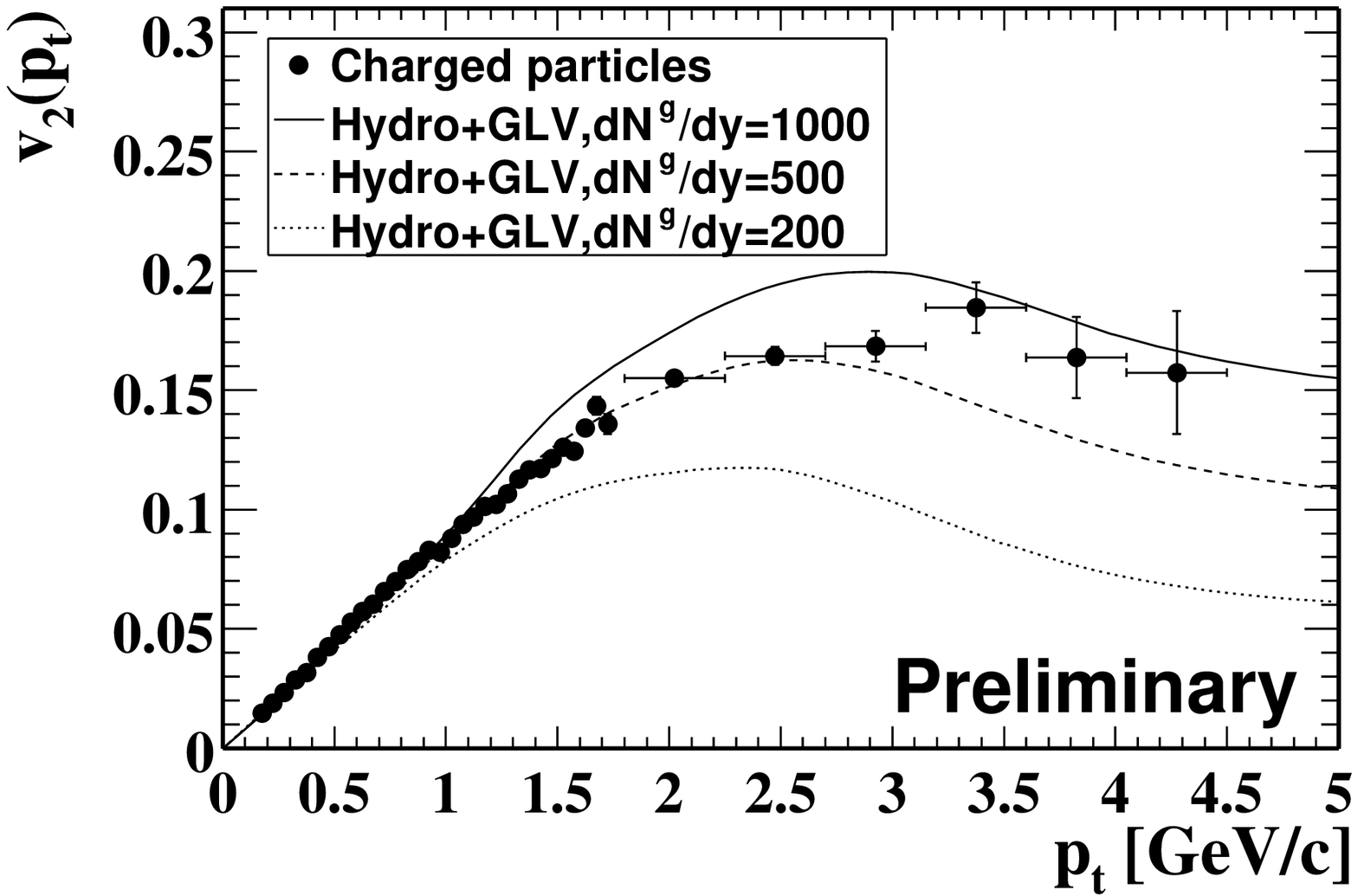}
    \caption{(Left Panel) $v_2$($p_t$) for charged particles and minimum-bias events,
      compared to hydrodynamic calculations.
             (Right Panel) $v_2$($p_t$) for charged particles and minimum-bias events,
      compared to pQCD calculations with different assumptions about the initial gluon density.}
    \label{highpt_pqcd}
  \end{minipage}
  \label{fig:STAR_flow}
\end{figure}

\begin{figure}
  \includegraphics[height=.3\textheight]{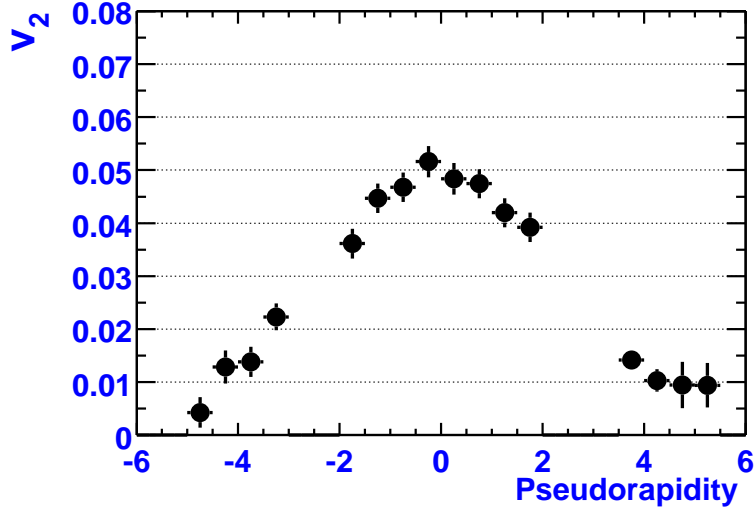}
  \caption{PHOBOS result on the elliptic flow $v_2$ as a function of pseudorapidity.}
\label{fig:phobos_flow_eta}
\end{figure}

The STAR experiment has published \cite{star-flow} results on $v_{2}$ as a function of collision
centrality.  Preliminary results from PHOBOS \cite{phobos-flow-qm} 
show reasonable agreement, while preliminary results
from PHENIX \cite{phenix-flow-qm} are somewhat larger though 
with a different transverse momentum selection.
The PHOBOS experiment has measured the 
elliptic flow $v_2$ as a function of pseudorapidity as shown in Fig.~\ref{fig:phobos_flow_eta}.
Preliminary results from STAR~\cite{star-flow2} 
as a function of transverse momentum for minimum bias collisions are shown
in Fig.~\ref{fig:STAR_flow}.  In the left panel, STAR results are shown compared with a
hydrodynamic model calculation which shows excellent agreement up to $p_{T} \approx 1.5$ GeV.
This agreement may indicate a large degree of thermalization in the early stages, thus
validating the assumption of hydrodynamic expansion.  In a specific model 
calculation \cite{kolb},
they estimate an initial thermal energy density $\epsilon=20~GeV/fm^{3}$ at a time 
$\tau=0.6~fm/c$.  The disagreement between the hydrodynamic calculation at large transverse
momentum has been attributed in some models to the high $p_T$ hadrons being the result
of jet fragmentation.  Normally jets would be uncorrelated with the reaction plane, but if
there is energy loss, as discussed in the next section, the path traveled by the parton
in the medium depends on its relative orientation to the reaction plane.  
In the right panel, the sensitivity of the high $p_T$ elliptic flow to
the initial gluon density is shown \cite{ivan}.  
These measurements and other preliminary results on flow for different particle species 
provide further tests to these models.

There are other preliminary results on particle momentum distributions and two particle
correlations that indicate large
transverse expansion of the system, described as radial flow.  Theoreticians are currently working
towards a complete description of many of these observables.  In particle the preliminary
two particle correlation (HBT) results from STAR are a challenge to include into a full
hydrodynamic picture and thus represent a critical measurement.

\subsection{Hard Process Probes of the Plasma}

An ideal experiment would be to contain the quark-gluon plasma and
then send well calibrated probes through it and measure the
resulting transparency or opacity of the system.  Despite the
great challenges heavy ion collisions present, there are such
calibrated probes but they must be self-generated in the collision
itself. An excellent probe is a parton or quark-antiquark pair.  A quark traversing a
color confined medium of hadrons sees a relatively transparent
system.  However, a parton passing through a hot color deconfined
medium will lose substantial energy via gluon 
radiation. In fact, because the radiated gluons can couple to each
other, the energy loss is proportional to the square of the path
length traversed \cite{mueller,mueller2}.

\begin{figure}
  \includegraphics[height=.5\textheight]{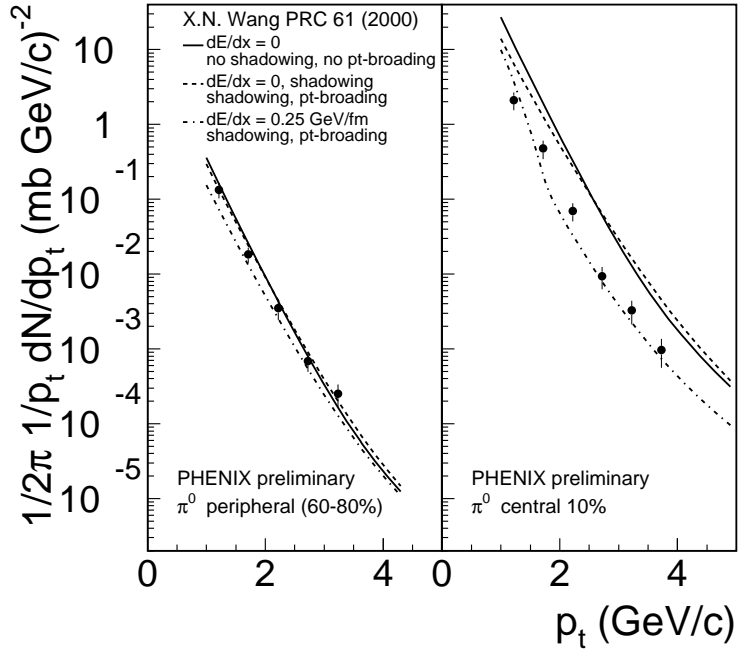}
  \caption{Preliminary PHENIX invariant multiplicity of identified $\pi^{0}$ as a function
of transverse momentum are shown for peripheral and central collisions.  Comparison with
theoretical calculations with and without parton energy loss are also shown.}
   \label{fig:phx-pizero}
\end{figure}

The source of these partons is from hard processes producing
back-to-back parton jets.  In a deconfined medium the parton will
lose energy before escaping the system and fragmenting into a jet
cone of hadrons.  The total energy of the initial parton jet is
conserved since eventually the radiated gluons will also
hadronize.  It is likely that the radiated gluons will have a
larger angular dispersion than the normally measured jet cone.
Thus one might be able to measure a modification in the apparent
jet shape.  A more striking signature is that since when the parton
fragments into hadrons it has less energy, the fragmentation will
result in a much reduced energy for the leading hadron.  Thus, a
measurement of high transverse momentum hadrons ($\pi^{0},
\pi^{+/-}, K^{+/-}, h^{+/-}$) is a strong indicator of the opacity of the
medium.

The PHENIX experiment has measured the distribution of identified $\pi^{0}$ for
both central and peripheral Au+Au collisions as shown in Fig.~\ref{fig:phx-pizero}~\cite{phx-qm}.
The peripheral results appear to be in good agreement within systematic errors
of an extrapolation from $pp$ collisions scaled up by the number of binary collisions
expected in this centrality class.  However, the central collision results show
a significant suppression in the $\pi^{0}$ yield relative to this point like scaling 
expected for large momentum transfer parton-parton interactions.  
If the created fireball in RHIC collisions is
transparent to quark jets, then we expect the yield of high
$p_{T}$ hadrons to obey point-like scaling and equal the $pp$ (or
equivalently $p\overline{p}$) distribution scaled up by the number of
binary $NN$ collisions, or equivalently by the nuclear
thickness function $T_{AA}$.  This is not what is observed.
A more sophisticated
calculation \cite{wang-prc} yields the same qualitative conclusion.

\begin{figure}
  \includegraphics[height=.5\textheight]{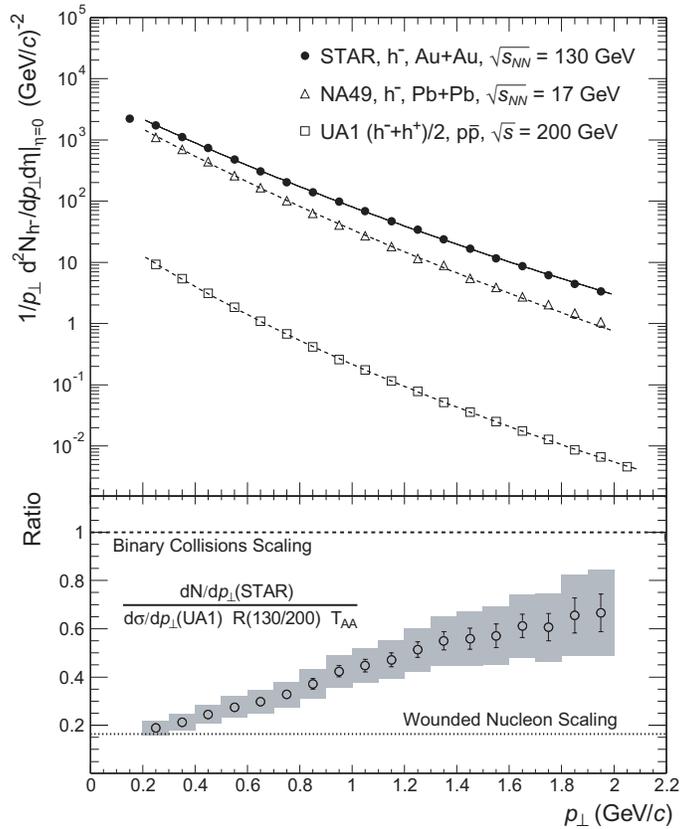}
  \caption{Invariant multiplicity of unidentified charged hadrons as a function of transverse momentum (top).
Ratio of unidentified charge hadrons per calculated binary
collision from Au + Au central collisions to those from p +
p($\overline{p}$) collisions extrapolated to $\sqrt{s}=130~GeV$ as
a function of transverse momentum (bottom).} \label{fig:star_pt}
\end{figure}

The STAR experiment has recently submitted for publication~\cite{star-highpt} the
$p_{T}$ spectra for unidentified negatively charged hadrons in
central Au + Au collisions as shown in Fig.~\ref{fig:star_pt}.
Also shown are the equivalent spectra from experiment NA49 at the
CERN-SPS at $\sqrt{s}=17~GeV$ and from UA1 in $p\overline{p}$ at
$\sqrt{s}=200~GeV$.  The STAR spectra is then divided by the spectra
from $p\overline{p}$ scaled by the number of binary collisions, and
the result is shown
in the lower panel of Fig~\ref{fig:star_pt}.  At low transverse
momentum the particle production is dominated by soft interactions
which scale with the number of wounded nucleons as indicated by
the line at 0.2.  The rise from 0.2 as a function of $p_{T}$ certainly has
a large contribution from hydrodynamic flow that will push particles to
higher transverse momentum in central Au+Au collisions.

\begin{figure}
  \includegraphics[height=.4\textheight]{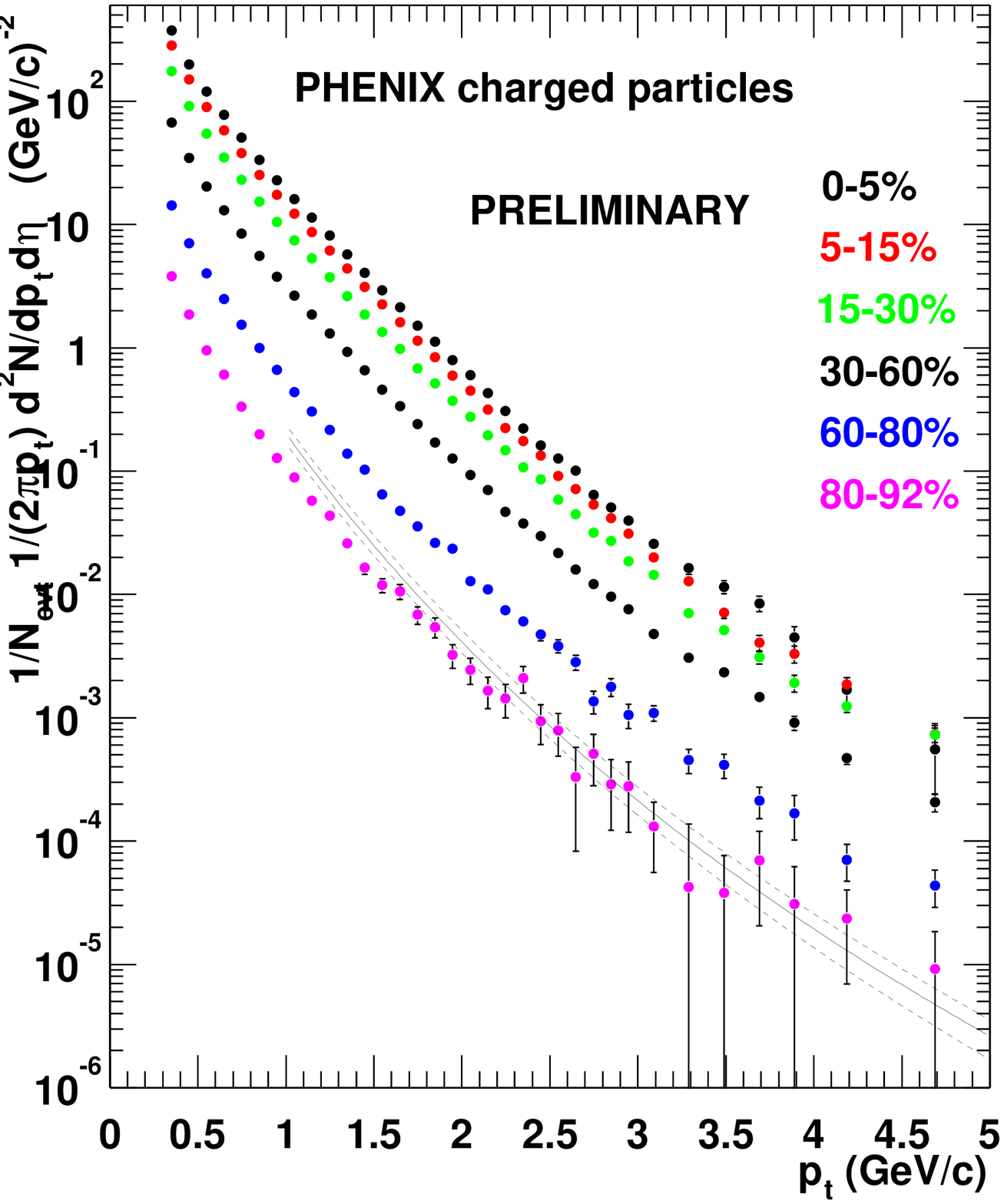}
  \caption{PHENIX preliminary results for unidentified charged hadron invariant
multiplicity as a function of transverse momentum.}
  \label{fig:phx-spectra}
\end{figure}

\begin{figure}
  \includegraphics[height=.3\textheight]{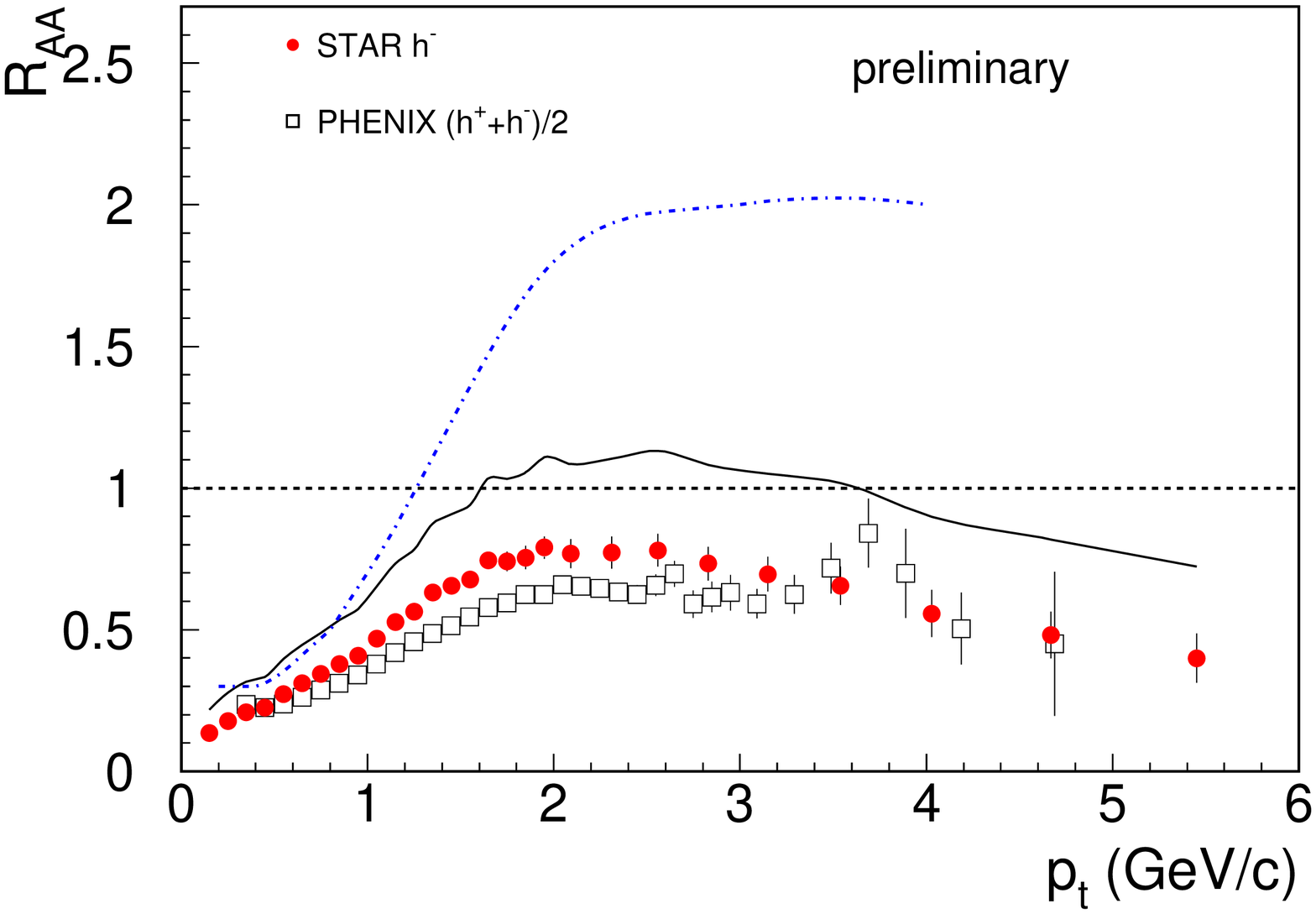}
  \caption{Ratio of unidentified charge hadrons per calculated binary
collision from Au + Au central collisions to those from p + p($\overline{p}$)
collisions (extrapolated to $\sqrt{s}=130~GeV$)
as a function of transverse momentum (GeV).  The solid line is the systematic
error band on the ratio.  The dashed line is the average result from experiments
at the lower energy CERN-SPS.}
  \label{fig:highpt_ratio}
\end{figure}

The PHENIX experiment has shown preliminary results extending out further in 
transverse momentum.  Preliminary results for six centrality classes are shown
from PHENIX in Fig~\ref{fig:phx-spectra}.  STAR also has preliminary results
for central collisions extending out to $p_{T}>5$ GeV that are in reasonable agreement
with the PHENIX results.  If one takes the ratio of the central spectra to the
unidentified spectra in $pp$ collisions scaled up by the number of binary collisions
one gets a ratio $R_{AA}$ as shown in Fig~\ref{fig:highpt_ratio}.
It needs to be noted that there is no $pp$ data at $\sqrt{s}=130$ GeV and thus an extrapolation
to that energy is done to calculate $R_{AA}$.  This extrapolation is included in the
systematic error band, and should be reduced when both experiments measure the
spectra in $pp$ in Run II.  

There are many important physics points to understand in these results.  The ratio
appears to stay below one, although that is a marginal conclusion with the present
systematic errors.  However, this is certainly in qualitative agreement with a parton
energy loss scenario, as also seen in the observed suppression in the PHENIX $\pi^{0}$ spectra.  
In contrast, the CERN-SPS results show an enhancement that has been attributed to the Cronin effect, or
initial state parton scattering that gives a $k_T$ kick to the final transverse
momentum distribution.  This expected enhancement makes the suppression seen at RHIC all
the more striking.

There are a number of open questions that must be considered before drawing any conclusions.  
The most basic is that these $p_{T}$ values are low relative to where one might have
confidence in the applicability of perturbative QCD calculations.  In addition, the separation
between soft and hard scale physics is blurred in this $p_{T}$ range, and in fact the 
CERN-SPS ratio $R_{AA}$ has also been explained in terms of hydrodynamic boosting of the
soft physics to higher $p_T$.  The preliminary results from PHENIX on the ratios of 
$\pi/K/p(\overline{p})$ in the middle of this $p_T$ range look more like soft physics than
a parton fragmentation function in vacuum.  One additional point of concern is that these
models of energy loss assume that the parton exits the collision region before finally
fragmenting into a jet of forward hadrons.  Thus the final hadronization takes place in vacuum.  
In the $p_T$ range of these early measurements, that conclusion is not so clear.  The parton
is traveling through the medium with various $k_T$ scatters, and if it hadronizes
inside a bath of other particles, the leading hadrons may be slowed down by inelastic collisions
with co-moving pions.  Lastly, the point-like scaling is known to be violated due to the
nuclear shadowing of parton distribution functions.  These nuclear modifications are known to 
reduce the pdf for quarks of order 20\% for $x\approx10^{-2}$; however, the shadowing for
gluons is not currently measured.  The calculations of \cite{wang-prc} have included modeling of this
shadowing, but must be viewed with caution at this time.  
These points need further theoretical investigation.  In addition, as will be 
discussed in the next section, many of these concerns are reduced when the measurements
extend to much higher transverse momentum.

\section{Run II at RHIC}

Run II at RHIC is currently starting and the expectation is to have ten weeks of running Au + Au
at the full energy $\sqrt{s}=200$ GeV per nucleon pair at the full design luminosity.  
This should allow experiments
to sample physics from well over one billion AuAu collisions.  After these ten weeks, there will
be the first polarized proton run, which represents the beginning of the RHIC spin program. In
addition it provides crucial comparison data sets for the heavy ion program.  
Beyond this time, the program
is not yet determined, but there is active discussion about lighter ion collisions, lower energy
running, and possibly deuteron-A collision studies.

In terms of initial conditions, lighter ion running and running at lower energies will be crucial
for drawing more firm conclusions about parton saturation and understanding the initial energy density.
The thermalization question will be addressed further by the measurement of direct photons and
lepton pairs, and chemical freeze-out can be tested with more excited state resonance measurements
and multistrange baryons.  Two particle correlation measurements have already been
crucial tests of the space-time evolution of the system, and possible measurements as a function
of the reaction plane will be an important correlated measurement to elliptic flow.

Probably the area most enhanced by the high luminosity running will the the probes of the plasma.
In the Au + Au running, we expect to see hadron spectra about beyond $p_{T} > 10$ GeV which helps to
address many of the current concerns with the preliminary Run I results.  In addition, more identified
hadrons will be measured at high transverse momentum.  Another key observable will be back-to-back
correlations at high $p_T$ in order to better calibrated possible parton energy loss scenarios. 
In addition, a second category of probes via heavy $q\overline{q}$ pairs that form vector mesons
such as $J/\psi,\psi',\Upsilon$ will be measured.  Associated measurements of charm and bottom
production via single and correlated lepton pairs will be important.
All of these measurements are eagerly anticipated.

\section{Looking Even Further Ahead}
The RHIC program has a long lifetime ahead of it.  The first glimpse at the physics from Run I
has proven to be very exciting and interesting.  There are clearly a number of systematic studies
that will be necessary to untangle competing effects and physics.  A full energy scan for
excitation function measurements, varying the collision geometry and an extensive comparison
proton-proton and proton-nucleus program is called for.  
Understanding the nature of the QCD vacuum at high temperatures is the primary goal for the field,
and we are on a good start towards that goal.

\begin{theacknowledgments}
JLN would like to thank the conference organizers for the opportunity to
present these first results on behalf of the entire RHIC community.  In
addition, useful conversations and material from S. Kelly, B. Jacak, P. Steinberg, 
T. Ullrich, C. Zhang are acknowledged.
\end{theacknowledgments}

% choose bibtex style depending on layout style and options used in
% sample:

\doingARLO[\bibliographystyle{aipproc}]
          {\ifthenelse{\equal{\AIPcitestyleselect}{num}}
             {\bibliographystyle{arlonum}}
             {\bibliographystyle{arlobib}}
          }
\bibliography{nagle}

\end{document}